\documentclass{ws-procs975x65}            
\begin{document}
\title{CP violation measurements at the LHCb experiment}

\author{L. Pescatore$^*$, on behalf of the LHCb collaboration}

\address{$^*$University of Birmingham, Birmingham, UK\\
$^*$E-mail: luca.pescatore@cern.ch}



\begin{abstract}
Decays of b-hadrons are the ideal place to perform measurements of CP violation. Many decay channels allow
 to over-constrain the unitarity triangles of the CKM matrix and test the SM hypothesis that a single
  phase is the origin of all CP violation. Charm decays also allow for tests of the SM. Recent results from LHCb are reviewed.
\end{abstract}


\bodymatter
\section{Introduction}

CP violation was first observed by Cronin and
Fitch\cite{Christenson:1964fg} in 1964 in the kaon sector and is by
now well established in the Standard Model (SM).
The LHCb experiment is performing precision measurements in order to consolidate the consistency of the CKM picture and
 look for deviations from the CP Violation (CPV) expected in the SM.
A selection of recent results from the LHCb experiment are presented. Where not explicitly stated all results are based
on the analysis of 1~fb$^{-1}$ of data collected in 2011 at a
proton-proton collision energy of 7~TeV.

\section{Measurement of the $\gamma$ angle}

One of the angles of the unitarity triangle is $\gamma = arg\left[ -V_{ud}V_{ub}^*/V_{cd}V_{cb}^* \right]$.
At the moment it has the weakest experimental constrains and therefore its measurement is
an important test of the CKM consistency. LHCb expects to achieve a precision of 7$^{\circ}$,
after the analysis of the 3~fb$^{-1}$ collected in 2011 and 2012 has been finished.
The angle is measured using $B\rightarrow Dh$ decays, where $h$ can be a pion or a kaon.
In these decays $\gamma$ arises from the interference of $b\rightarrow u$ and  $b\rightarrow c$ transitions.

The measurement can be carried out using D decays in CP eigenstates $KK$ and
$\pi\pi$ (GLW method\cite{Gronau:1990ra}), or also in $K\pi$ (ADS method\cite{Atwood:1996ci}).
In the latter case the $B^-\rightarrow D^0 K^-$ decay is colour favoured but the
$D^0$ decay in $K^+\pi^-$ is CKM suppressed, yielding large interference.
Combining the two methods and including also $D\rightarrow K_Shh$ decays
the value of $\gamma$ is measured to be $(62 \pm
12)^{\circ}$\cite{LHCb-CONF-2013-006}.
Most of the analysis included are on 1~fb$^{-1}$ of data and being
updated to 3~fb$^{-1}$.

\section{$B_s$ mixing and $\phi_s$ measurement}

In the neutral B system, mixing is possible thanks to weak interaction box diagrams.
The angle $\phi_s$ arises from the interference of $B_s$ decays with and without mixing.
The value of $\phi_s$ is well known in the SM, $\phi_s = -0.0364 \pm 0.0016$~rad\cite{Charles:2011va}, but New Physics (NP)
could add large phases. Here a measurement of $\phi_s$  is presented using
$B_s \rightarrow J/\psi\pi\pi$ decays, which has recently been updated
using 3~fb$^{-1}$ of data\cite{Aaij:2014siy}.
$\phi_s$ is measured from a time-dependent amplitude analysis, namely fitting the decay-time distributions
of $B_s$ events that are flavour tagged at production. The fit considers the effect of a finite time resolution
(40.3~fs in LHCb) and a decay-time dependent acceptance.
The angle $\phi_s$ is measured to be $0.070 \pm 0.068 \text{(stat)} \pm 0.008 \text{(sys)}$~rad. This is the single
most precise measurement to date and it is in agreement with the SM.
The measurement was also performed constraining the fit to zero CP violation and the value found is
$0.075 \pm 0.065 \text{ (stat) } \pm 0.008 \text{ (sys) }$~rad\cite{Aaij:2014dka}.

\section{Direct CP violation}

In this section a few of the latest measurements of direct CPV at LHCb
in the B and D sector are presented.
In this type of analysis we look for a difference in the decay rates of charge conjugate decays.
\begin{equation}
A_{raw} = \frac{ N(I \rightarrow f) - N(\overline{I} \rightarrow \overline{f}) }
{ N(I \rightarrow f) + N(\overline{I} \rightarrow \overline{f})}
\end{equation}
In general, the true CP asymmetry, $A_{CP}$, can be written as $A_{CP} =
A_{raw} + A_{det} + A_{prod}$, where $A_{det}$ is the bias due to a different detection efficiency for particles and antiparticles.
This can be partly due to differences in performance in the left and right part of the detector and also to the
fact that the nuclear interaction with the detector material is different for particles and antiparticles.
To reduce the first effect LHCb's magnet polarity is periodically reversed so that left-right differences
are averaged out. Finally, $A_{prod}$ is the asymmetry arising from D
or B meson production effects.

\subsection{CP asymmetry in $B^0\rightarrow \phi K^{*0}$}

In this analysis the quantity $\Delta A_{CP} = A_{CP}(B^0\rightarrow
\phi K^{*0}) - A_{CP}(B^0\rightarrow J/\psi K^{*0})$ is measured.
The channel $J/\psi K^*$, which has same final daughters as the signal, is used as a control channel
in order to cancel detection and production asymmetries.
The result is $\Delta A_{CP} = (1.5 \pm 3.2 \text{(stat)} \pm  0.5
\text{(sys)})\%$\cite{Aaij:2014tpa}.

\subsection{3-body charmless decays}

$B^\pm\rightarrow \pi^-\pi^+\pi^\pm$ and $B^\pm\rightarrow K^-K^+\pi^\pm$
decays are analysed looking for direct CPV\cite{Aaij:2013bla}. In a
previous paper\cite{Aaij:2013sfa} $B\rightarrow K^\pm \pi^-\pi^+$ and
$B\rightarrow K^\pm K^-K^+$ were also considered.
In this case the detection asymmetry of the pion is measured in LHCb using a tag-and-probe method.
The detection asymmetry of kaons is measured using $D$ decays to $KK$ and $K\pi$ and
correcting for the pion asymmetry. Finally, the production asymmetry is measured using
$B^\pm \rightarrow J/\psi K^\pm$ decays, where the CP asymmetry is assumed to be zero, and correcting for
the detection asymmetry of the kaon.

The results\cite{Aaij:2013bla}~\cite{Aaij:2013sfa} are:
\begin{align}
\nonumber
 &A_{CP}(B^\pm \rightarrow K^\pm\pi^-\pi^+) = 0.032 \pm 0.008 \text{(stat)} \pm 0.004 \text{(sys)} \pm 0.007 \text{ ($J/ψK^\pm$) }
\nonumber
\\ &A_{CP}(B^\pm \rightarrow K^\pm K^-K^+) = -0.043 \pm 0.009 \text{(stat)} \pm 0.003 \text{(sys)} \pm 0.007 \text{ ($J/ψK^\pm$) }
\nonumber
\\ &A_{CP}(B^\pm \rightarrow \pi^\pm K^-K^+) = -0.141 \pm 0.040
\text{(stat)} \pm 0.0018 \text{(sys)} \pm 0.007 \text{ ($J/ψK^\pm$) }
\nonumber
\\ &A_{CP}(B^\pm \rightarrow \pi^\pm\pi^-\pi^+) = 0.117 \pm 0.021
\text{(stat)} \pm 0.009 \text{(sys)} \pm 0.007 \text{ ($J/ψK^\pm$) }
\nonumber
\end{align}
\begin{figure}
\includegraphics[width=0.54\textwidth]{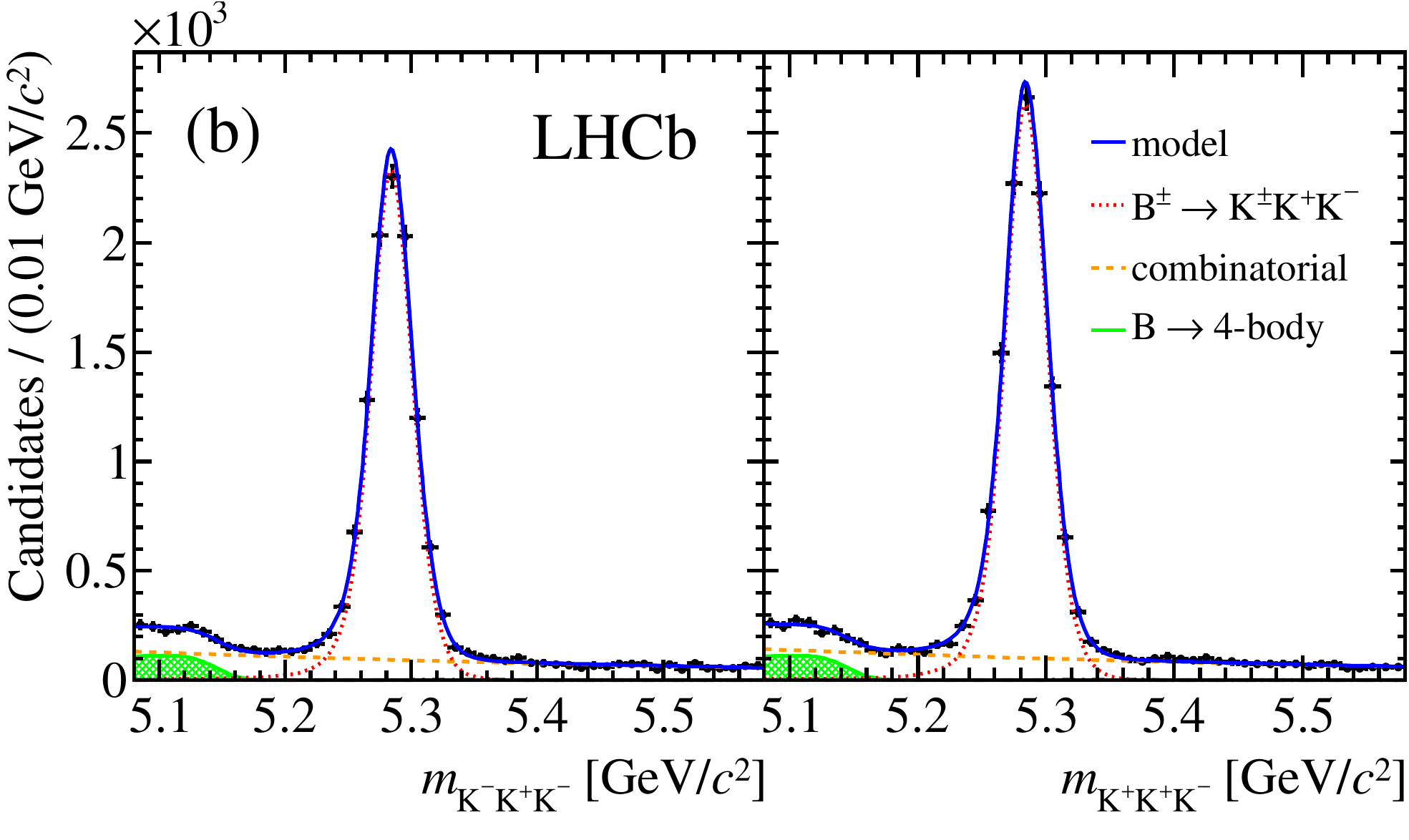}
\includegraphics[width=0.45\textwidth]{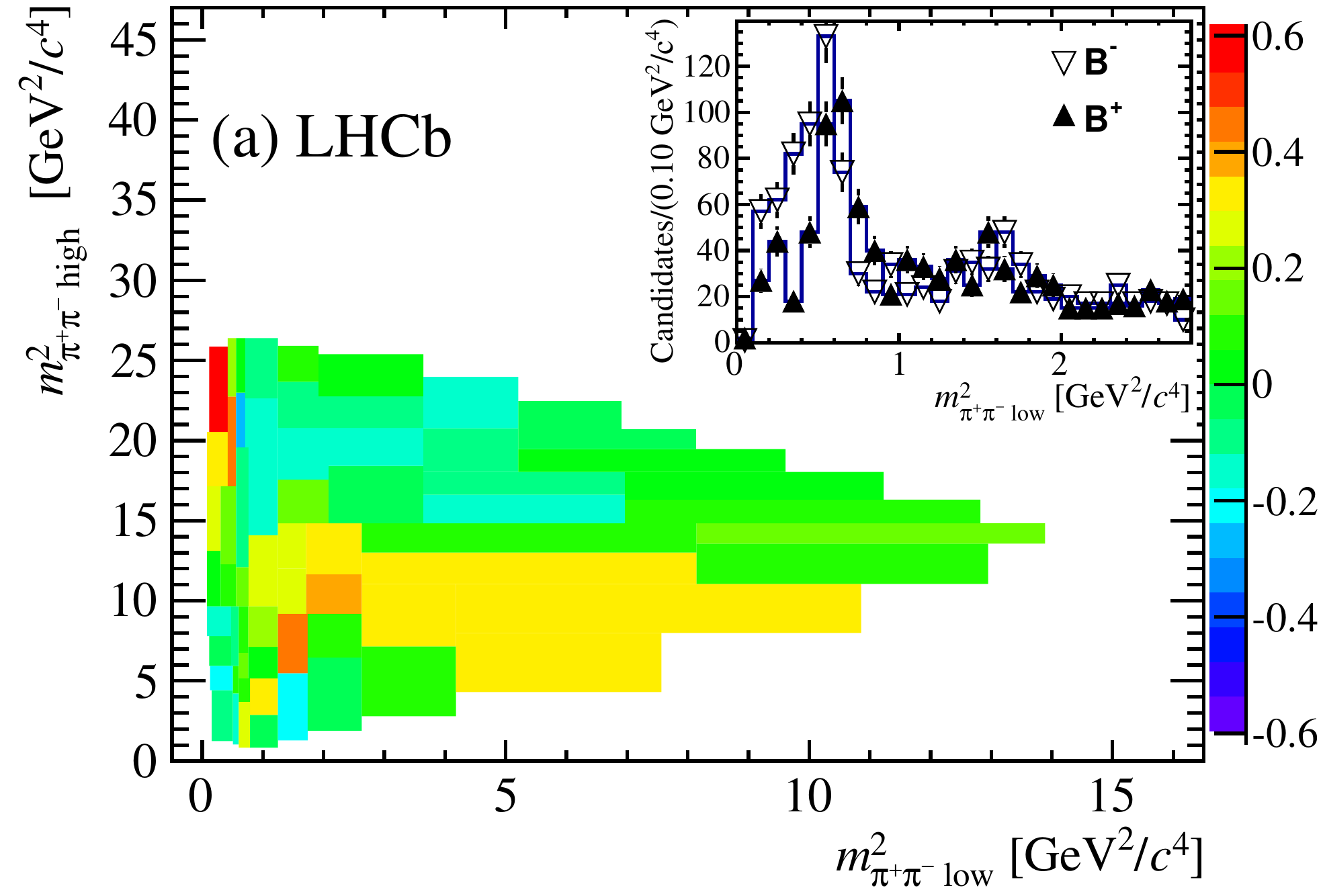}
\caption{Invariant masses of $K^+K^+K^-$ and $K^-K^+K^-$ systems
  (left) and raw asymmetries in the
  $B^\pm\rightarrow\pi^\pm\pi^+\pi^-$ Dalitz plot (right).}
\label{fig:3body}
\end{figure}

The significance of $A_{CP}(B^\pm \rightarrow K^\pm K^+K^-)$ exceeds three standard deviations, yielding the first evidence
of a CP asymmetry in charmless three-body decays. Figure
\ref{fig:3body} (left) shows $K^+KK$ and $K^-KK$ invariant masses.
The Dalitz plots were also studied, looking for localised
asymmetries. Figure \ref{fig:3body} (right) shows asymmetries in the
raw number of events in bins of the Dalitz plot for $B^\pm\rightarrow\pi^\pm\pi^+\pi^-$ decays.
Large asymmetries are clearly visible in the localised region of phase space defined
 by $m^2_{\pi^+\pi^-_{high}} > 15$~GeV$^2/$c$^4$ and
 $m^2_{\pi^+\pi^-_{low}} < 0.4$~GeV$^2/$c$^4$, where no significant
 contributions from resonances are expected.

\subsection{CP asymmetry in $\Lambda_b \rightarrow J/\psi p\pi^-$}

\begin{figure}
\centering
\includegraphics[width=0.38\textwidth]{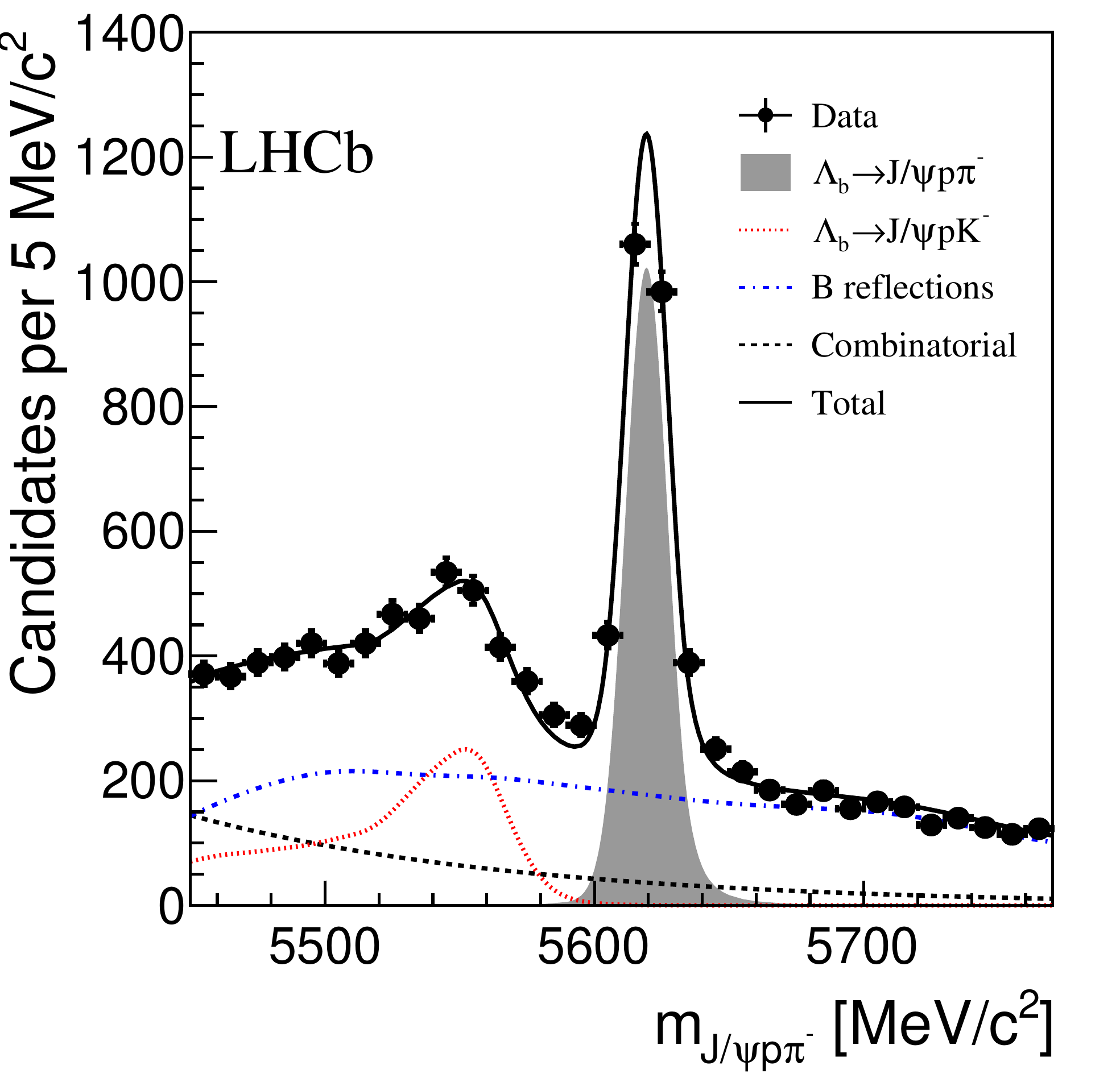}
\includegraphics[width=0.38\textwidth]{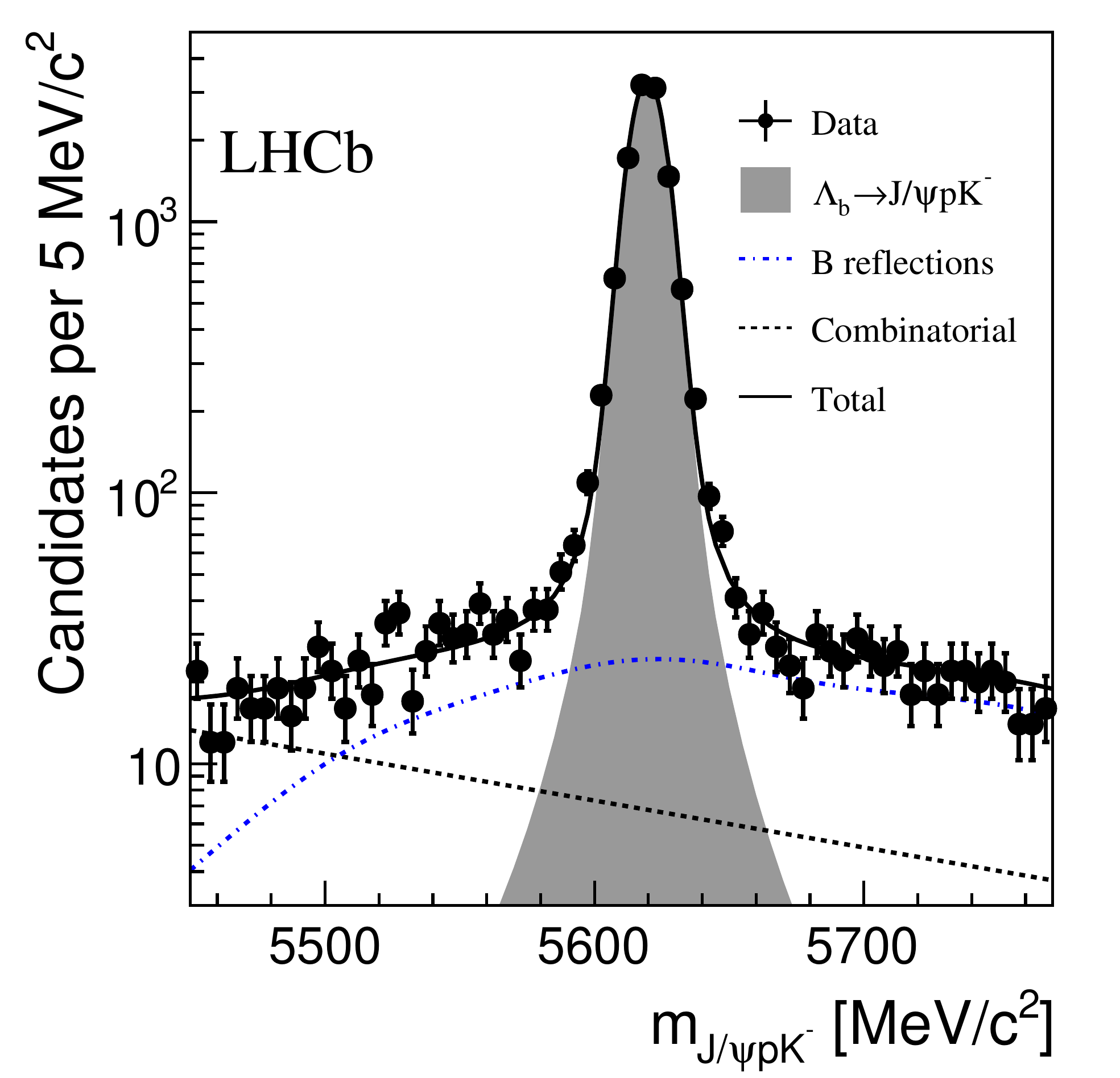}
\caption{Invariant masses of the $J/\psi p\pi^-$ (left) and $J/\psi
  pK^-$ (right) systems.}
\label{fig:LbACP}
\end{figure}

In this analysis the $\Delta A_{CP}$ between the Cabibbo suppressed (CS) $\Lambda_b \rightarrow J/\psi p\pi^-$
and Cabibbo favoured (CF) $\Lambda_b \rightarrow J/\psi pK^-$ is measured. The data sample is split in two exclusive samples
using the particle ID information on the meson in the final state. In Fig.~\ref{fig:LbACP}
invariant masses for the two samples are reported.
The production asymmetry is the same for the two channels and cancels in the difference.
Therefore $\Delta A_{CP} = A_{CP}(p\pi) - A_{CP}(pK) + A_{det}(\pi) - A_{det}(K)$.
The detection asymmetries of kaon and pion can be determined using the $B^0\rightarrow J/\psi K^{*0}$
decay, where the CP asymmetry is assumed to be zero and the production
asymmetry is measured to be consistent with zero\cite{LHCb:2012kz},
giving $A_{CP}(B^0\rightarrow J/\psi K^{*0}) = A_{det}(\pi) - A_{det}(K)$.
The result, using 3~fb$^{-1}$ of data, is $\Delta A_{CP} = (5.7 \pm 2.3 \pm 1.2)\%$ with a 2.2
$\sigma$ deviation from zero\cite{Aaij:2014zoa}.

\subsection{Direct CPV in $D^0\rightarrow h^+h^-$}

This analysis uses $D^0$ decays in CP eigenstates ($KK$ and $\pi\pi$). The D flavour
is tagged at production by looking at the charge of the muon in the decay $B\rightarrow D\mu\nu X$.
These are Cabibbo suppressed decays where penguin diagrams may
contribute allowing space for New Physics.
Furthermore, using charm decays, we have $O(10^6)$ events available,
a factor 1000 more than B decays, which allows more precise measurements.
The asymmetry is expected to be equal in magnitude and opposite in sign so the quantity measured
is the $A_{CP}$ difference  $\Delta A_{CP} = A_{CP}(KK) - A_{CP}(\pi\pi)$, where detection and production asymmetries cancel.
On 3~fb$^{-1}$ of data, this is measured to be $0.14 \pm 0.16\text{(stat)} \pm 0.08\text{(sys)} $, in agreement with the SM\cite{Aaij:2014gsa}.

\subsection{Direct CPV in $D^+\rightarrow \pi^-\pi^+\pi^+$}

\begin{figure}
\centering
\includegraphics[width=0.43\textwidth]{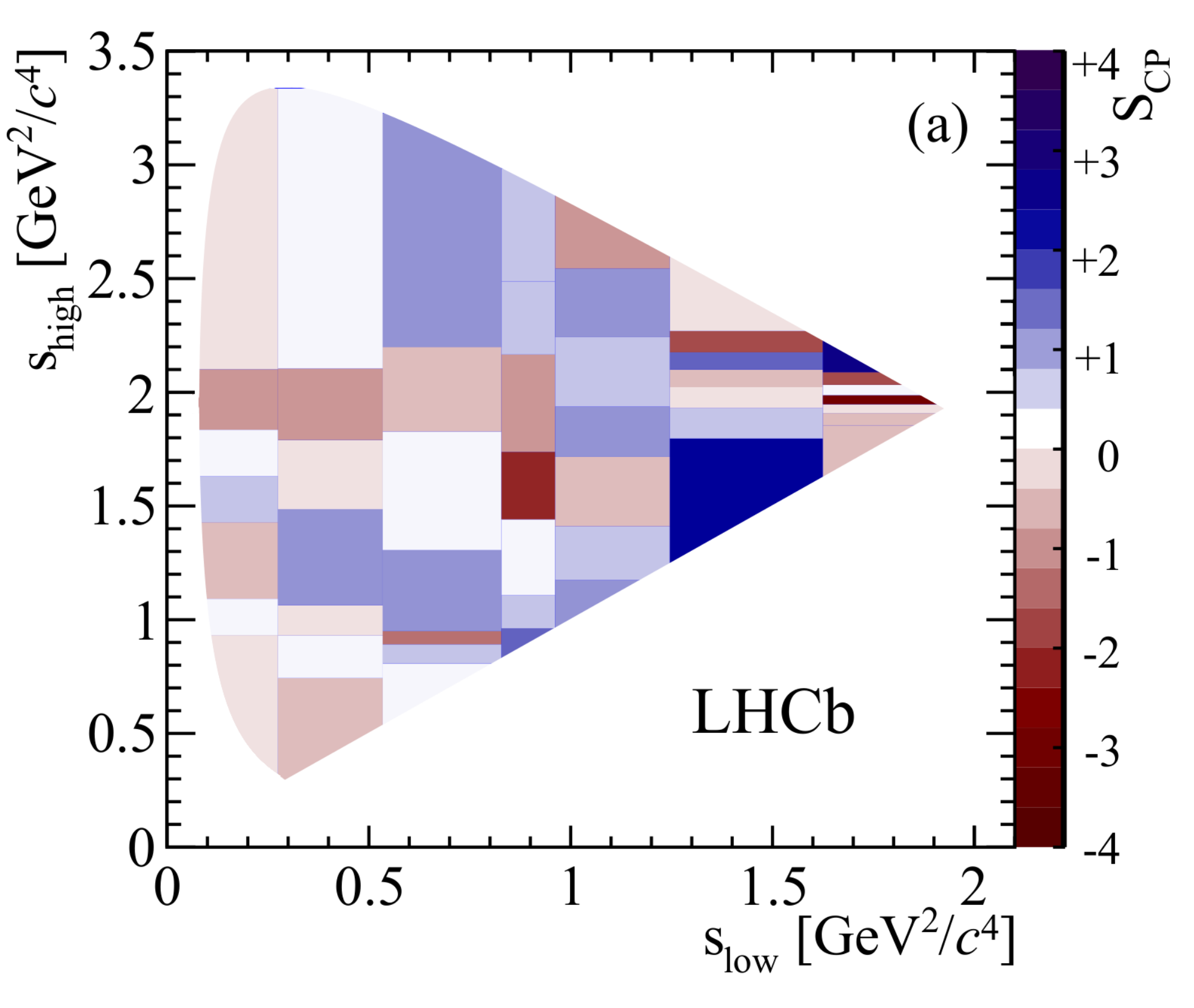}
\includegraphics[width=0.43\textwidth]{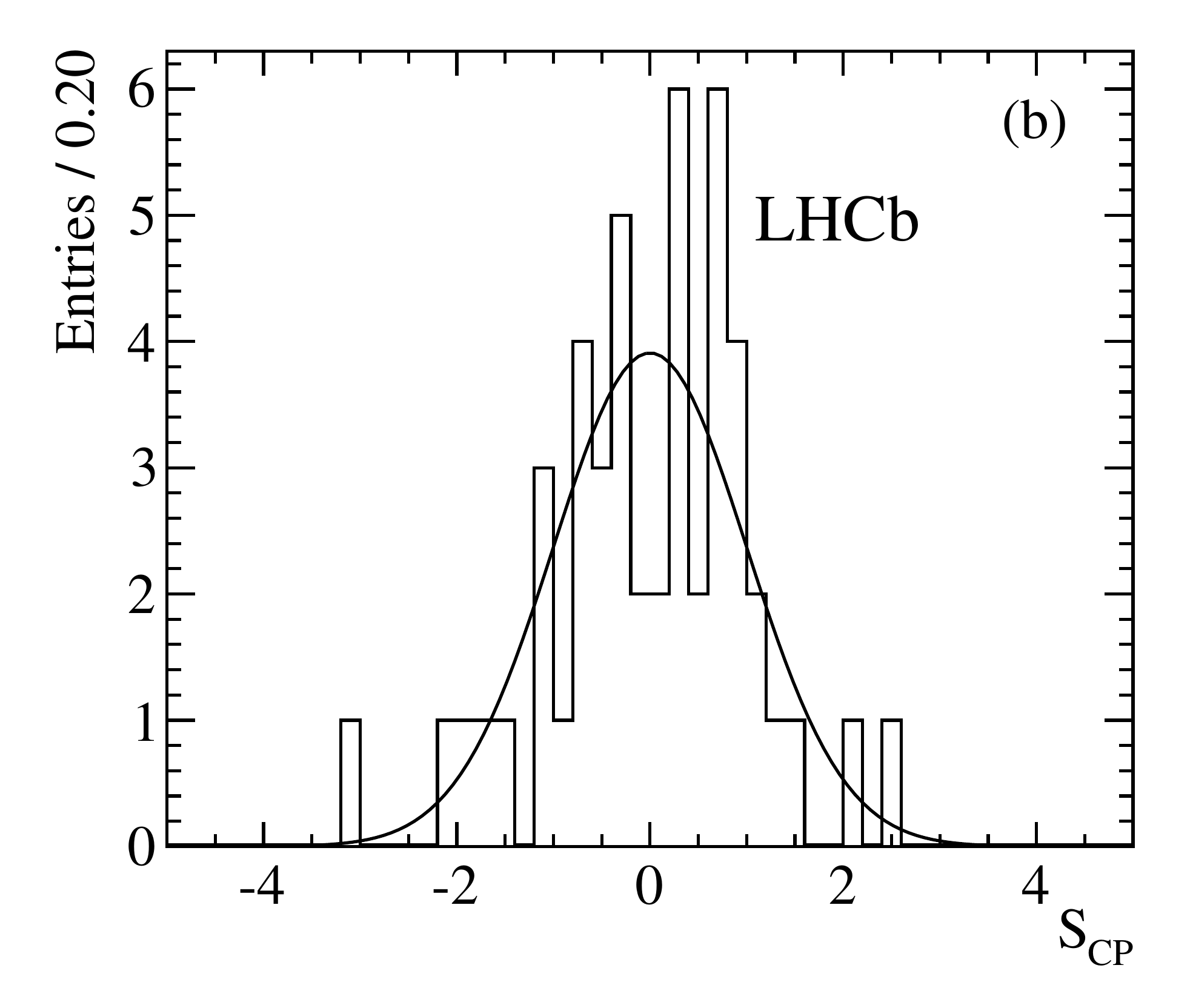}
\caption{Significances in the Dalitz plot (left) and their
  distribution (right).}
\label{fig:D3pi}
\end{figure}

In this analysis the Dalitz plot of the Cabibbo suppressed $D^+\rightarrow \pi^-\pi^+\pi^+$ decay is studied, looking for local CPV.
The analysis is performed by dividing the Dalitz plot into bins and calculating in each bin a significance function defined as:
\begin{equation}
S_i \equiv \frac{ N_i(D^+) - \alpha N_i(D^-) }{ \sqrt{ \alpha
    (N_i(D^+) + N_i(D^-)) } },
\end{equation}
where $\alpha = \sum N_i(D^+) / \sum N_i(D^-) = 0.992\pm0.001$ accounts for global asymmetries.
Figure \ref{fig:D3pi} shows the significances in the Dalitz plot and how
the values are distributed.
If there is no local CPV the significances should follow a Gaussian distribution.
Finally, a $\chi^2$ test is used to find a p-value for the non-CPV hypothesis. No local CPV is found\cite{Aaij:2013jxa}.

\section{Conclusions}

The most recent combination of the CKM angle $\gamma$ measurements at
LHCb was presented, where the value found is $(62\pm12)^{\circ}$.
A number of results of CPV searches in $B_s$ mixing and decay rates
were also performed where in general a good
agreement with the SM is found.
No evidence for CPV in charm decays is found.

\bibliographystyle{ws-procs975x65}
\bibliography{references}

\end{document}